\documentclass[pra, superscriptaddress, twocolumn, showpacs]{revtex4}

\usepackage{amsfonts}
\usepackage{amssymb}
\usepackage[dvips]{graphicx}
\usepackage{amsmath}

\newcommand{\bra}[1]    {\langle #1|}
\newcommand{\ket}[1]    {| #1 \rangle}
\newcommand{\braket}[2]{\langle #1 | #2 \rangle}
\newcommand{\e}         {\mathrm{e}}

\newcommand{\Tr}{\textrm{Tr}}
\renewcommand{\t}[1]{\textrm{#1}}

\begin{document}

\title{Quantum phase estimation with lossy interferometers}



\author{R.~Demkowicz-Dobrzanski}
\affiliation{Institute of Physics, Nicolaus Copernicus University, Grudziadzka 5, PL-87-100 Toru\'{n}, Poland}
\author{U.~Dorner}
\affiliation{Clarendon Laboratory, University of Oxford, Parks Road, Oxford OX1 3PU, United Kingdom}
\author{B.~J.~Smith}
\affiliation{Clarendon Laboratory, University of Oxford, Parks Road, Oxford OX1 3PU, United Kingdom}
\affiliation{Centre for Quantum Technologies, National University of Singapore, 117543 Singapore, Singapore}
\author{J.~S.~Lundeen}
\affiliation{Clarendon Laboratory, University of Oxford, Parks Road, Oxford OX1 3PU, United Kingdom}
\author{W.~Wasilewski}
\affiliation{Institute of Experimental Physics, University of Warsaw, Ho\.{z}a 69, PL-00-681 Warsaw, Poland}
\author{K.~Banaszek}
\affiliation{Institute of Physics, Nicolaus Copernicus University, Grudziadzka 5, PL-87-100 Toru\'{n}, Poland}
\author{I.~A.~Walmsley}
\affiliation{Clarendon Laboratory, University of Oxford, Parks Road, Oxford OX1 3PU, United Kingdom}
\date{\today}

\begin{abstract}
  We give a detailed discussion of optimal quantum states for optical
  two-mode interferometry in the presence of photon losses.  We derive
  analytical formulae for the precision of phase estimation obtainable
  using quantum states of light with a definite photon number and prove
  that maximization of the precision is a convex optimization problem.
  The corresponding optimal precision, i.e. the lowest possible
  uncertainty, is shown to beat the standard quantum limit thus
  outperforming classical interferometry.
  Furthermore, we
  discuss more general inputs: states with
  indefinite photon number and states with photons distributed between
  distinguishable time bins. We prove that neither of these is
  helpful in improving phase estimation precision.
\end{abstract}

\maketitle

\section{Introduction}

The strong sensitivity of certain quantum states to small variations
of external parameters opens up great opportunities for devising
high-precision measurements, e.g. of length and time, with
unprecedented accuracy. A particularly important physical measurement
technique is interferometry. Its numerous variations include Ramsey
spectroscopy in atomic physics, optical interferometry in
gravitational wave detectors, laser gyroscopes and optical imaging to
name but a few.  Understanding limits on its performance in realistic
situations under given resources is therefore of fundamental
importance to metrology.  In this paper we examine the fundamental
limits of the precision of optical interferometry in the presence of
photon losses for quantum states of light with definite photon number.

Optical interferometry aims to estimate the relative phase of two
modes, or two ``arms'', of the interferometer.  This estimation
process requires a certain amount of resources which is typically
identified to be the number of photons, $N$, used for the measurement.
The best precision which can be obtained using classical states of
light scales like $1/\sqrt{N}$, the so-called standard quantum limit
(SQL). Using non-classical states of light this precision can be greatly
improved, ideally leading to Heisenberg-limited scaling, $1/N$
\cite{Giovannetti2004,Giovannetti2006}. Indeed, recent years have seen
many experimental proof-of-principle demonstrations of beating the SQL
using quantum strategies in various interferometric setups
~\cite{Mitchell2004,Eisenberg2005,Nagata2007,Resch2007,Higgins2007,Higgins2008}.
Unfortunately, highly non-classical states of light which potentially lead
to Heisenberg-limited sensitivity are very fragile with
respect to unwanted but unavoidable noise in experiments. In quantum-enhanced optical
interferometry, the loss of photons is the most common
and potentially the most devastating type of noise that one encounters.
In particular it was noted that highly entangled quantum states,
optimal for interferometry in the lossless case -- N00N states
\cite{Bollinger1996} -- are extremely fragile.  Even for moderate
losses they are outperformed by purely classical
states~\cite{Huelga1997,Shaji2007,Sarovar2006,Rubin2007,Gilbert2008,Huver2008,Dorner2008}.
 A different approach has been taken in \cite{Meiser2008}, where the noise
arising from imperfect preparation of a state has been investigated.

In~\cite{Dorner2008}, the first systematic approach was taken in order
to determine the structure of optical states optimal for
interferometry in the presence of losses. The best possible precision
using input states with definite photon number was given. In this
paper we elaborate and extend the ideas presented
in~\cite{Dorner2008}. Our treatment is based on general quantum
measurement
theory~\cite{Helstrom1976,Holevo1982,Braunstein1994,Braunstein1996}.
The quantity of interest is the lowest possible uncertainty attainable in parameter
estimation, inversely proportional to the square root of the
quantum Fisher information~\cite{Braunstein1994}. The quantum Fisher
information depends only on the state of the system and not on the
measurement procedure. We show that the
optimization of the quantum Fisher information can be done effectively over the class of input states
 which have a definite photon number. Since these input
states are subject to unavoidable photon losses they will degrade into
mixed states and their suitability for phase estimation is
compromised. Our optimization takes this into account yielding the
most suitable input states in the presence of photon losses leading to
the highest possible quantum Fisher information, and hence to the
best possible precision.  We give a detailed description of the
noise model and calculate an analytic expression for the quantum
Fisher information.  The latter is shown to be a concave function on a
convex set and therefore suited for efficient convex optimization
methods. We numerically determine the optimal input states, compare
them to alternative quantum and classical strategies, and show that
they can beat the SQL. We note that the corresponding precision, which
lies between the SQL and the Heisenberg limit (depending on the loss
rates), defines the best possible precision for optical two-mode
interferometry. In addition to this we discuss a measurement
procedure that allows one to achieve the optimal precision, in terms of
a positive operator-valued measure (POVM), and the possibility of
using states with indefinite photon number or distinguishable photons.
We show that neither of these generalizations improves the estimation
precision, and consequently the state with definite photon number and
indistinguishable photons are optimal.

The paper is organized as follows. In Sec.~\ref{sec:phasestimation} a
general scheme of quantum phase estimation is presented, and the
notion of optimality is defined. In Sec.~\ref{sec:fisher} we introduce
the quantum Fisher information and discuss its most important
properties. In Sec.~\ref{sec:interferometry} we derive an explicit
formula for the quantum Fisher information and prove that it is a
concave function of input state parameters. In
Sec.~\ref{sec:optimalstates} we discuss the structure of the optimal
states for interferometry and compare them to alternative strategies
and states.
In Sec.~\ref{sec:measurement} we discuss a measurement
with which it is possible to achieve optimal precision.
Finally, in
Sec.~\ref{sec:generalstates} we discuss possible generalizations of
the considered quantum states, particularly states with indefinite
photon number and the case when photons are distinguishable.

\section{Phase estimation}
\label{sec:phasestimation}
\begin{figure}[t]
  \centering\includegraphics[width=8cm]{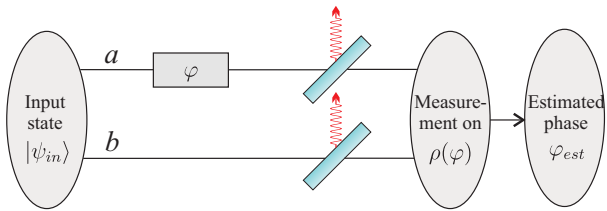}
\caption{%
  Interferometric phase estimation scheme. An input state is fed into
  an interferometer consisting of two channels. Channel $a$ acquires a
  phase $\varphi$ relative to channel $b$. Measurements are performed
  on the output state yielding an estimated value, $\varphi_{est}$, of
  the phase $\varphi$.  The beam splitters symbolize photon losses.}
\label{fig1}
\end{figure}
We consider a general interferometer with two arms as shown in
Fig.~\ref{fig1}. A pure input state $\ket{\psi_{in}}$ is fed into the
interferometer and acquires a phase $\varphi$ in the channel $a$ relative
to the channel $b$. Both channels, or ``arms'', of the interferometer are
subject to photon losses which can be modelled by fictitious beam
splitters inserted at arbitrary locations in both channels. The
output of the interferometer therefore needs to be described in general by a
mixed state $\rho(\varphi)$.  A measurement, represented by a positive
operator valued measure (POVM) $\{\Pi_i\}$ which defines a probability
distribution for the measurement outcomes,
\begin{equation}
p(i|\varphi) = \Tr\left[\Pi_i \rho(\varphi)\right],
\end{equation}
is subsequently performed on the output state $\rho(\varphi)$. An
estimated value of the true phase $\varphi$ is obtained by applying an
estimator that assigns to a particular measurement result $i$ an estimated value $\varphi_{est}(i)$. We aim to estimate the phase
$\varphi$ as precisely as possible.

Two more elements need to specified in order to make the problem of
finding the optimal phase estimation strategy well defined: an a
priori knowledge on the phase distribution $p(\varphi)$ and a cost
function $C(\varphi,\varphi_{est})$ which can be seen as a measure for
the uncertainty in the estimated phase.  For phase estimation, the
optimal choice of the input state, measurement and estimator is the
one that minimizes the average cost function
\begin{equation}
\label{eq:cost}
\langle C\rangle=\int \t{d}\varphi\, p(\varphi) \sum_i p(i|\varphi) C(\varphi,\varphi_{est}(i)).
\end{equation}
At this point two different approaches are most often pursued. In
the \emph{global approach} one assumes initial ignorance about the
actual value of $\varphi$, which corresponds to the choice
$p(\varphi)=1/2\pi$. The solution of the problem then yields an
estimation strategy which performs equally well irrespectively of the
actual value of the estimated phase.  In the \emph{local approach}, on
the other hand, the assumption is that the value of the actual phase
lies in the vicinity of a known phase $\varphi_0$. More precisely, the
a priori probability is chosen to be $p(\varphi)=\delta(\varphi -
\varphi_0)$, while the estimator is required to be \emph{locally
  unbiased} \footnote{Sometimes a weaker condition of \emph{asymptotic
    local unbiasedness} is imposed i.e. unbiasedness of estimation
  when the estimator is calculated using data coming from a very large
  number of repetitions of an experiment (see
  e.g.~\cite{Braunstein1994}).}
\begin{equation} \left. \frac{\t{d}}{\t{d} \varphi } \sum_i
  p(i|\varphi) \varphi_{est}(i) \right|_{\varphi=\varphi_0} = 1.
\end{equation}
The above condition is equivalent to a statement that the estimator
will on average yield the true value of $\varphi$ up to the first
order in $(\varphi-\varphi_0)$. Notice that without local unbiasedness
the estimation problem would be trivial (and also useless)
since in order to minimize $\langle C\rangle$ in Eq.~(\ref{eq:cost}), with
$p(\varphi)=\delta(\varphi - \varphi_0)$, the optimal choice for the
estimator would be simply $\varphi_{est}(i) = \varphi_0$, and the
choice of the measurement would be irrelevant.

The local approach is useful when we are interested in small
deviations of the phase from a known one. A significant advantage of
the local approach over the global one is that for a natural choice of a quadratic cost function, there exist explicit
lower bounds on $\langle C\rangle$ based on the Fisher information
(see Sec.~\ref{sec:fisher}). In many practical situations these bounds
are tight and the optimization over the measurement and the estimator
can be avoided. Moreover, the local approach may also be useful in
situations when there is no a priori knowledge on the phase $\varphi$.
If many copies of a state are given, one can first perform a rough
measurement (even not optimal) on a small fraction of copies in order
to narrow down the range of potential values of $\varphi$ so that they
lie in a vicinity of a known phase $\varphi_0$, and then perform the
optimal estimation using the local approach. This strategy will yield
a high accuracy estimation for the global approach, without the need
of optimizing the measurement and the estimator, since the rough
measurement performed on small fraction of copies, even if not
optimal, will not significantly influence the final accuracy
\cite{gill2000}.

\section{Fisher information}
\label{sec:fisher}
In what follows we take the local approach. Since in this case we
deal with small deviations of estimated phases from the true one,
it is natural to choose
$C(\varphi,\varphi^\prime)=(\varphi-\varphi^\prime)^2$ as the cost
function. Our goal is to find the optimal state, measurement and
locally unbiased estimator minimizing the expression
\begin{equation}
\label{eq:variance}
(\delta \varphi)^2 = \sum_i p(i|\varphi)[\varphi_{est}(i)- \varphi]^2.
\end{equation}
To minimize the standard deviation $\delta\varphi$ given by
the above formula we use an upper bound on
$\delta\varphi$ based on the Fisher information
\cite{Fisher1925,Helstrom1976}.  For a given POVM $\{ \Pi_i \}$
and state $\rho(\varphi)$ defining the probabilities $p(i|\varphi)$, the
Cram\'er-Rao inequality bounds the variance that can be
obtained using any locally unbiased estimator,
\begin{equation}
\label{eq:cramerclassical}
(\delta\varphi)^2 \geq \frac{1}{F},
\end{equation}
where the Fisher information $F$ is given by
\begin{equation}
 \quad F=\sum_i \frac{1}{p(i|\varphi)} \left(\frac{\partial p(i|\varphi)}{\partial \varphi}\right)^2.
\label{eq:fisherinfo}
\end{equation}
If an experiment is repeated $\nu$ times the bound reads
\begin{equation}
\label{eq:cramer_repeat}
(\delta\varphi)^2 \geq \frac{1}{\nu F}.
\end{equation}
For large $\nu$ the Cram\'er-Rao bound is asymptotically achieved by
the maximum likelihood
estimator~\cite{Fisher1925,Helstrom1976,Braunstein1994}.

Optimization over the measurements yields the quantum Cram\'er-Rao
bound~\cite{Helstrom1976,Holevo1982,Braunstein1994,Braunstein1996}
\begin{equation}
\label{eq:qcr}
(\delta \varphi)^2 \geq \frac{1}{F}  \geq \frac{1}{F_Q},
\end{equation}
where the quantum Fisher information $F_Q$ is given by
\begin{equation}
\quad F_Q=\Tr[ \rho(\varphi) A^2].
\end{equation}
The Hermitian operator $A$ is called the ``symmetric logarithmic
derivative'' (SLD) and is implicitly defined via the relation
\begin{equation}
\frac{\partial \rho(\varphi)}{\partial \varphi} = \frac{1}{2}[ A \rho(\varphi) + \rho(\varphi) A].
\end{equation}
In the eigenbasis of $\rho(\varphi)$, $A$ is given by
\begin{equation}
\label{eq:sld}
(A)_{ij} = \frac{2}{p_i + p_j}  \left[ \rho^\prime(\varphi) \right]_{ij},
\end{equation}
where $\rho^\prime(\varphi)= \frac{\partial \rho(\varphi)}{\partial
  \varphi}$ and the $p_i$ are the eigenvalues of $\rho(\varphi)$ (whenever
$p_i+p_j=0$ we set $(A)_{ij}=0$).  It has been shown that a
measurement saturating the quantum Cram\'er-Rao bound exists and is
given by a projective measurement on the eigenbasis of
$A$~\cite{Braunstein1994,Braunstein1996}.

For the sake of completeness we state some important properties of
$F_Q$ (see e.g.~\cite{Helstrom1976,Fujiwara2001}):

(i) Let $\rho(\varphi)$, $\sigma(\varphi)$ be two density
    matrices supported on orthogonal subspaces,
    $\mathcal{S}_{\rho(\varphi)} \bot \mathcal{S}_{\sigma(\varphi)}$,
    which do not cease to be orthogonal for an infinitesimal change of
    $\varphi$, i.e.  $(\mathcal{S}_{\rho(\varphi)}\bigcup
    \mathcal{S}_{\rho^\prime(\varphi)}) \bot
    (\mathcal{S}_{\sigma(\varphi)}\bigcup
    \mathcal{S}_{\sigma^\prime(\varphi)})$, then $F_Q$ is linear on
    the direct sum
\begin{align}
\label{eq:lindir}
F_Q[p\rho(\varphi) \oplus (1-p)&\sigma(\varphi)] \nonumber\\
&=pF_Q[\rho(\varphi)] + (1-p)F_Q[\sigma(\varphi)].
\end{align}
(ii) $F_Q$ is convex
\begin{align}
\label{eq:convexity}
F_Q[p\rho(\varphi) + (1-p)&\sigma(\varphi)] \nonumber\\
& \leq pF_Q[\rho(\varphi)] + (1-p)F_Q[\sigma(\varphi)]
\end{align}
(iii) For pure states
    $\rho(\varphi)=\ket{\psi(\varphi)}\bra{\psi(\varphi)}$, $F_Q$ reads
\begin{equation}
\label{eq:fishpure}
F_Q=4[\braket{\psi^\prime(\varphi)}{\psi^\prime(\varphi)} - |\braket{\psi^\prime(\varphi)}{\psi(\varphi)}|^2],
\end{equation}
where $\ket{\psi^\prime(\varphi)}=\partial \ket{\psi(\varphi)}/ \partial \varphi$.

Property (i) is due to the fact that the SLD for $p\rho(\varphi)
\oplus (1-p)\sigma(\varphi)$ is a direct product of SLDs for
$p\rho(\varphi)$ and $(1-p)\sigma(\varphi)$, respectively. Property (ii) is a
consequence of the fact that by (i) the right hand side of (ii) can be
viewed as the quantum Fisher information of the state $p \rho(\varphi) \otimes \ket{0}\bra{0} +
(1-p)\sigma(\varphi)\otimes \ket{1}\bra{1}$, where $\ket{0}$, $\ket{1}$
are orthogonal ancillary states, while the left hand side is $F_Q$ of the state
after tracing out the ancillary system. Furthermore, $F_Q$ is non-increasing under stochastic operations~\cite{petz1996}
(tracing out the ancilla is an example).
Property (iii) is a consequence of Eqs.~(\ref{eq:qcr},\ref{eq:sld}),
since for a pure state
 \begin{equation}
 A=2 (\ket{\psi(\varphi)}\bra{\psi^\prime(\varphi)} + \ket{\psi^\prime(\varphi)}\bra{\psi(\varphi)}).
 \end{equation}
 The measurement saturating the quantum Cram\'er-Rao bound in this
 case is a von Neumann measurement projecting on any orthonormal basis
 containing two vectors
\begin{equation}
\ket{e_{\pm}}= \frac{1}{\sqrt{2}}\left( \ket{\psi(\varphi)} \pm \ket{\psi^{\prime \bot}(\varphi)} \right),
\end{equation}
where
\begin{equation}
\ket{\psi^{\prime \bot}(\varphi)}= \frac{1}{\mathcal{N}}(\ket{\psi^\prime(\varphi)} - \braket{\psi(\varphi)}{\psi^\prime(\varphi)}\ket{\psi(\varphi)} )
\end{equation}
is the normalized vector orthogonal to $\ket{\psi(\varphi)}$ lying in
the space spanned by $\ket{\psi(\varphi)}$ and
$\ket{\psi^\prime(\varphi)}$.

\section{Interferometry with losses}
\label{sec:interferometry}
Assuming that we have $N$ photons at our disposal, we aim to find the
input state that allows performing phase estimation with the best
precision possible, i.e. yielding the highest value of the quantum
Fisher information $F_Q$. In particular we consider the most general
pure two-mode input state with definite photon number $N$,
\begin{equation}
  \ket{\psi_{in}} = \sum_{k=0}^N \alpha_k \ket{k,N-k},
\label{eq:input}
\end{equation}
where $\ket{k,N-k}$ abbreviates the Fock state $\ket{k}_a\ket{N-k}_b$.
This class of states includes the N00N state which, in the absence of
losses, leads to Heisenberg limited precision, but is very fragile in
the presence of noise. We are therefore looking for states which lead
possibly to a lower precision than the N00N state, but which are more
robust with respect to photon losses.  Moreover, although states of
the form~(\ref{eq:input}) seem to be a restriction, we show in
Sec.~\ref{sec:generalstates} that our treatment effectively includes
states with indefinite photon number.

In the following subsections we show how states of the
form~(\ref{eq:input}) are influenced by photon losses, calculate its
quantum Fisher information and show that the latter can be maximized
(thus minimizing $\delta\varphi$) by means of convex optimization
methods.

\subsection{Noise model}
Losses are modeled by fictitious beam splitters of transmissivity
$\eta_a$, $\eta_b$ in channels $a$ and $b$ respectively, and cause a
Fock state $\ket{k,N-k}$ to evolve into
\begin{equation}
  \ket{k,N-k} \mapsto \sum_{l_a = 0}^k\sum_{l_b=0}^{N-k} \sqrt{B_{l_a l_b}^k} \ket{k-l_a, N-k -l_b}\otimes \ket{l_a,l_b},
\end{equation}
where $\ket{l_a, l_b}$ represents the state of two ancillary modes
carrying $l_a$ and $l_b$ photons lost from modes $a$ and $b$
respectively, while
\begin{equation}
B^k_{l_al_b}= \binom{k}{l_a}\binom{N-k}{l_b}\eta_a^{k}(\eta_a^{-1}-1)^{l_a} \eta_b^{N-k}(\eta_b^{-1}-1)^{l_b}.
\end{equation}
Including the phase accumulation $\ket{k,N-k}\mapsto e^{i k\varphi} \ket{k,N-k}$ and tracing out the ancillary modes results in the output density matrix
\begin{equation}
\label{eq:rhoout}
\rho(\varphi)=\sum_{l_a=0}^N \sum_{l_b=0}^{N-l_a} p_{l_a l_b} \ket{\xi_{l_a l_b}(\varphi)} \bra{\xi_{l_a l_b}(\varphi)},
\end{equation}
where
\begin{equation}
  \ket{\xi_{l_a l_b}(\varphi)}=\frac{1}{\sqrt{p_{l_a l_b}}}\sum_{k =
    l_a}^{N-l_b} \alpha_k e^{i k \varphi } \sqrt{B_{l_a l_b}^k
  }\ket{k-l_a,N-k-l_b}
\end{equation}
is the conditional pure state corresponding to the event when $l_a$
and $l_b$ photons are lost in modes $a$ and $b$ respectively, and
$p_{l_a l_b}$ is the normalization factor corresponding to the probability of that
event.

Equivalently, the loss process can be described by a master equation
for two independently damped harmonic oscillators with loss rates
$\gamma_{a,b} = |\ln\eta_{a,b}|/t$, where $t$ is time, the solution of
which is given by
\begin{equation}
\rho = \sum_{k,l=0}^\infty K_{l,a} K_{k,b}\rho_{in} K_{k,b}^\dagger K_{l,a}^\dagger
\label{eq:kraus}
\end{equation}
with Kraus operators
\begin{equation}
K_{l,a} = (1-\eta_a)^\frac{l}{2}\eta_a^{\frac{1}{2}{\hat a}^\dagger \hat a} {\hat a}^l/\sqrt{l!},
\end{equation}
where $\hat a$ is the annihilation operator for mode $a$, and
analogously for mode $b$. This state acquires a phase through the
transformation $\rho(\varphi) = \e^{- i \varphi \hat a^\dagger \hat a}
\rho \e^{ i \varphi \hat a^\dagger \hat a}$.
Notice that thanks to the relation
\begin{equation}
\hat{a}^l e^{-i  \varphi \hat a^\dagger \hat a} = e^{ -i  \varphi \hat a^\dagger \hat a} \hat{a}^l e^{i \varphi l}
\end{equation}
we can commute the phase operator with the Kraus operators since the phase terms $e^{-i \varphi l}$ cancels out.
It is therefore irrelevant if photons are lost before, during or after channel $a$
acquires its relative phase with respect to $b$.

\subsection{Calculating the Fisher information}
\label{sec:calcfisher}
Using Eq.~(\ref{eq:rhoout}) for the output state, one can calculate
$F_Q$ with the help of Eqs.~(\ref{eq:qcr}) and~(\ref{eq:sld}). This
requires diagonalization of $\rho(\varphi)$ which can be carried out in the case of one-arm losses.
In the more general case of losses in both arms an
analytic calculation of $F_Q$ turns out to be infeasible. Nevertheless,
we are able to determine an upper bound to the quantum Fisher
information which, although not strictly tight for general input
states, is very close to $F_Q$ for the states we consider in Sec.~\ref{sec:optimalstates}.

\subsubsection{Losses in one arm}
We consider first the case $\eta_a=\eta$, $\eta_b=1$, i.e. when
losses are present in only one arm. As can be seen from Eq.~(\ref{eq:rhoout}),
in this case only states $\ket{\xi_{l_a l_b}(\varphi)}$ with
$l_b=0$ contribute to $\rho(\varphi)$.  Moreover, we have  $\braket{\xi_{l
    0}(\varphi)}{\xi_{l^\prime 0}(\varphi)}=\delta_{l l^\prime}$,
hence we can write the output state as a direct sum
\begin{equation}
  \rho(\varphi) = \bigoplus_{l=0}^N p_{l 0} \ket{\xi_{l 0}(\varphi)} \bra{\xi_{l 0}(\varphi)}.
\end{equation}
Making use of Eqs.~(\ref{eq:lindir}) and~(\ref{eq:fishpure}) we get a
formula for $F_Q$ with explicit dependence on the input state
parameters $\alpha_k$,
\begin{equation}
F_Q = 4\left(
\sum_{k=0}^N k^2 x_k - \sum_{l=0}^N\frac{\left( \sum_{k=l}^{N} x_k k B^k_{l0} \right)^2}{\sum_{k=l}^{N} x_k B^k_{l0}}
\right),
\label{eq:fisher1mode_a}
\end{equation}
where $x_k=|\alpha_k|^2$. In a more compact way the above
formula can be rewritten as
\begin{equation}
\label{eq:fisher1mode}
  F_Q=2\sum_{l=0}^N \frac{\boldsymbol{x}^T \mathbf{R}^{(l 0)} \boldsymbol{x}}{\boldsymbol{x}^T \boldsymbol{b}^{(l 0)}},
\end{equation}
where $\boldsymbol{x}$ is a vector containing variables $x_k$, while
the elements of the vector $\boldsymbol{b}^{(l_a l_b)}$ and the matrix
$\mathbf{R}^{(l_a l_b)}$ are given by
\begin{align}
  b^{(l_a l_b)}_k &=\left\{
\begin{array}{ccl}
    B^k_{l_a l_b} & \t{ if  } & l_a \leq k \leq N - l_b \\
    0  & \t{otherwise}
\end{array},
\right. \\
R^{(l_a l_b)}_{k,k^\prime}&=b^{(l_a l_b)}_k (k-k^\prime)^2
b^{(l_a l_b)}_{k^\prime}.
\end{align}

\subsubsection{Losses in two arms}\label{sec:calcfisherB}
If losses are present in both arms, then, in the most general case,
all $\ket{\xi_{l_a l_b}(\varphi)}$ contribute to $\rho(\varphi)$.
States with different total number of lost photons, $l=l_a + l_b$, are
still orthogonal. Using Eq.~(\ref{eq:lindir}) we can therefore write
\begin{equation}
  F_Q = \sum_{l=0}^N F_Q\left[\sum_{l_a=0}^{l} p_{l_a l-l_a} \ket{\xi_{l_a l-l_a}(\varphi)}\bra{\xi_{l_a l-l_a}(\varphi)}\right],
\end{equation}
where $F_Q[\cdot]$ denotes the quantum Fisher information of the state
in brackets.  Notice that states with the same $l$ are not necessarily
orthogonal. Consequently, the calculation of the above expression
requires solving an eigenvalue problem which is not feasible
analytically. Nevertheless, using the convexity of $F_Q$,
Eq.~(\ref{eq:convexity}), we obtain a bound
\begin{equation}
\label{eq:boundconvexity}
F_Q \leq \tilde{F}_Q = \sum_{l_a=0}^N \sum_{l_b=0}^{N-l_a} p_{l_a l_b} F_Q[\ket{\xi_{l_a l_b}(\varphi)} \bra{\xi_{l_a l_b}(\varphi)}],
\end{equation}
which can be calculated explicitly using Eq.~(\ref{eq:fishpure}). It
is given by
\begin{equation}
\tilde F_Q = 4\left(
\sum_{k=0}^N k^2 x_k - \sum_{l=0}^N\sum_{m=0}^{N-l}\frac{\left( \sum_{k=l}^{N-m} x_k k B^k_{lm} \right)^2}{\sum_{k=l}^{N-m} x_k B^k_{lm}}
\right),
\label{eq:fisherbound}
\end{equation}
or in a compact form,
\begin{equation}
\label{eq:fisher2mode}
  \tilde{F}_Q= 2\sum_{l_a=0}^N\sum_{l_b=0}^{N-l_a} \frac{\boldsymbol{x}^T \mathbf{R}^{(l_a l_b)} \boldsymbol{x}}{\boldsymbol{x}^T
  \boldsymbol{b}^{(l_a l_b)}}.
\end{equation}
When losses are present only in one arm (see previous paragraph)
$F_Q=\tilde{F}_Q$.  When losses are present in both arms, however, the
bound is not always tight. The difference $\tilde{F}_Q -F_Q$
originates from the non-orthogonality of $\ket{\xi_{l_a
    l_b}(\varphi)}$ for a fixed $l=l_a+l_b$, and physically
corresponds to lack of knowledge about how many photons were lost from
a particular mode. If this knowledge is not relevant then
$\tilde{F}_Q=F_Q$. This happens, e.g. in the case of the N00N state
$\alpha_0 \ket{N,0}+ \alpha_N \ket{0,N}$, when loss of even a single
photon renders the output states useless for phase estimation, hence
the knowledge of which mode the photons were lost does not influence
the value of $F_Q$.

\subsection{Concavity of Fisher information}
Even with the explicit formulae for $F_Q$ and $\tilde F_Q$ given by
Eqs.~(\ref{eq:fisher1mode}) and~(\ref{eq:fisher2mode}) it is in
general not possible to find an analytic solution for the optimal
state, i.e. values $x_k\geq 0$, $\sum_k x_k =1$ that maximize $F_Q$
(or $\tilde{F}_Q$).  However, we prove below that $\tilde{F}_Q$ is a
concave function of the $\{x_k\}$. Consequently, the problem amounts
to the maximization of a concave function $\tilde{F}_Q$ on a convex
set.  This allows for a feasible numerical constrained optimization using
interior-point method routines (e.g. implemented in \emph{Mathematica} 6.0), and more importantly, any
local maximum found is automatically the global maximum.

We prove concavity by showing that the Hessian $H_{ij}
=\frac{\partial^2 \tilde{F}_Q}{\partial x_i \partial x_j}$ is negative
semidefinite, i.e. for every vector $\boldsymbol{y}$ we have
$\boldsymbol{y}^T \mathbf{H} \boldsymbol{y} \leq 0$.  Using
Eq.~(\ref{eq:fisher2mode}) the Hessian $H_{ij}$ ($i,j \in
\{0,\dots,N\}$) reads
\begin{widetext}
\begin{equation}
  H_{ij} = 4 \sum_{k_1=0}^N \sum_{l_b=0}^{N-l_a} \frac{b^{(l_al_b)}_i
    b^{(l_al_b)}_j (\boldsymbol{x}^T
    \mathbf{R}^{(l_al_b)}\boldsymbol{x})- \boldsymbol{x}^T
    \boldsymbol{b}^{(l_al_b)}[b^{(l_al_b)}_i (\mathbf{R}^{(l_al_b)}
    \boldsymbol{x})_j + b^{(l_al_b)}_j(\mathbf{R}^{(l_al_b)}
    \boldsymbol{x})_i] +\boldsymbol{x}^T \boldsymbol{b}^{(l_al_b)}
    (\mathbf{R}^{(l_al_b)})_{i,j} \boldsymbol{x}^T
    \boldsymbol{b}^{(l_al_b)}}{(\boldsymbol{x}^T
    \boldsymbol{b}^{(l_al_b)})^3}.
\end{equation}
Since the denominator is always positive, it is sufficient to prove
that for every $l_a$, $l_b$ and for every vector $\boldsymbol{y}$ we have
\begin{equation}
(\boldsymbol{y}^T \boldsymbol{b}^{(l_al_b)})^2 (\boldsymbol{x}^T \mathbf{R}^{(l_al_b)}\boldsymbol{x})-
(\boldsymbol{x}^T   \boldsymbol{b}^{(l_al_b)} )(\boldsymbol{y}^T \boldsymbol{b}^{(l_al_b)}) (\boldsymbol{y}^T
\mathbf{R}^{(l_a,l_b)} \boldsymbol{x} + \boldsymbol{x}^T\mathbf{R}^{(l_a,l_b)}\boldsymbol{y})
+(\boldsymbol{x}^T   \boldsymbol{b}^{(l_al_b)})^2 \boldsymbol{y}^T \mathbf{R}^{(l_al_b)}  \boldsymbol{y} \leq 0.
\end{equation}
\end{widetext}
Introducing $K_{ij} = (i-j)^2$, $w_i = y_i b^{(l_al_b)}_i
(\boldsymbol{x}^T \boldsymbol{b}^{(l_al_b)})- x_i b^{(l_al_b)}_i
(\boldsymbol{y}^T \boldsymbol{b}^{(l_al_b)})$ the above condition can
be written equivalently as
\begin{equation}
\label{eq:concavitysimple}
\boldsymbol{w}^T \mathbf{K} \boldsymbol{w} \leq 0.
\end{equation}
Since $\sum_i w_i = 0$, it is sufficient to prove that $\mathbf{K}$ is
negative semi-definite on the set of vectors with coefficients summing
up to $0$.  To this end we define vectors $\boldsymbol{e}^{(i)}$
($i=1,\dots,N$), where $e_{0}^{(i)}=1$, $e^{(i)}_{i}=-1$,
$e^{(i)}_j=0$ (for $j \neq 0$, $j \neq i$), which span the space of
all vectors with coefficients summing up to zero.  Writing
$\boldsymbol{w}=\sum_{i=1}^N \beta_i \boldsymbol{e}^{(i)}$ and
noticing that $\boldsymbol{e}^{(i)T}\mathbf{K} \boldsymbol{e}^{(j)}= -2 i
j$ we arrive at
\begin{equation}
\boldsymbol{w}^T \mathbf{K} \boldsymbol{w} = -2 \sum_{i,j} \beta_i \beta_j i j = -2\left(\sum_i \beta_i i\right)^2 \leq 0,
\end{equation}
which proves that $\tilde{F}_Q$ is a concave function of the
$\{x_k\}$.

\section{Optimal states for phase estimation}
\label{sec:optimalstates}

In this section we discuss optimization results based on
Eqs.~(\ref{eq:fisher1mode_a}) and (\ref{eq:fisherbound}) derived in
the previous sections. The quantity we analyze is
$\delta\varphi_{min}\equiv 1/\sqrt{F_Q}$ corresponding to the best
possible precision for a fixed number of measurements $\nu$ (cf.
Eq.~(\ref{eq:qcr})). The only exception to this definition is in the
case of the optimal state for losses in both arms where we set
$\delta\varphi_{min}\equiv 1/\sqrt{\tilde F_Q}$ (see
Sec.~\ref{lossesintwoarms}). We compare the optimal precision to the
precision which is obtainable using various alternative states and
strategies.  In particular, we define the standard interferometric
limit (SIL) corresponding to the precision which can be achieved in a
classical reference experiment. This serves as a benchmark by which we
can judge the advantage of using quantum states of light over
classical ones.  For a given photon number $N$ the SIL is the
precision of phase estimation using a Mach-Zehnder interferometer in
which one arm gathers a phase $\varphi$, fed at one input port with a
coherent state $\ket{\alpha}$, where $|\alpha|^2=N$, and the vacuum at
the other port. The reflectivity of the two beam splitters can be
adjusted to achieve the best precision, while the measurement consists
of photon counting at the two output ports. Without any additional
reference beams the input coherent state should be regarded as a state
with unknown phase, and effectively described as a mixture of Fock
states with Poissonian statistics (see \cite{Molmer1997}, and the
discussion in Sec.~\ref{sec:indefiniteN}).  By Eq.~(\ref{eq:lindir}) the quantum Fisher information will be a weighted sum
of quantum Fisher information calculated for each input Fock state after it passes through the first beam splitter.
Taking the optimal value of transmissivity of the first beam splitter such that the ratio of the intensities in the arms $a$ and $b$ is $\sqrt{\eta_b/\eta_a}$ the final uncertainty in phase estimation achievable with this classical strategy is
given by
\begin{equation}
\delta\varphi_{min,\t{SIL}} = \frac{\sqrt{\eta_a} + \sqrt{\eta_b}} {2\sqrt{N\eta_a\eta_b}}.
\label{eq:sil}
\end{equation}
Note that the SIL scales in the
same way as the SQL, particularly for equal losses in both arms we
obtain $\delta\varphi_{\t{SIL}}=1/\sqrt{N\eta}$, where
$\eta=\eta_a=\eta_b$.

As a reference let us also calculate the minimum uncertainty achievable using $N$ photons prepared in an unbalanced N00N state:
\begin{equation}
\sqrt{x_N}\ket{N0} + \sqrt{x_0}\ket{0N}.
\label{eq:NOON_state}
\end{equation}
The quantum Cram{\'e}r-Rao bound yields:
\begin{equation}
\delta\varphi_{min,\t{N00N}} = \frac{ \eta_a^{N/2} + \eta_b^{N/2}} {2N \eta_a^{N/2}\eta_b^{N/2}},
\label{eq:noon_prec}
\end{equation}
where the optimal amplitudes are given by
\begin{equation}
x_0 = \frac{\eta_a^{N/2}}{\eta_a^{N/2} + \eta_b^{N/2}}
\end{equation}
and $x_N=1-x_0$.  Note that putting $N=1$ in Eq.~(\ref{eq:noon_prec})
coincides with Eq.~(\ref{eq:sil}). This implies that a coherent state performs equally well in phase estimation
as the equivalent number of single photons sent one-by-one.
In the following we will concentrate on the two
important scenarios of losses only in channel $a$ and equal losses in
both channels.

\subsection{Losses in one arm}
\begin{figure}
  \includegraphics[width=\columnwidth]{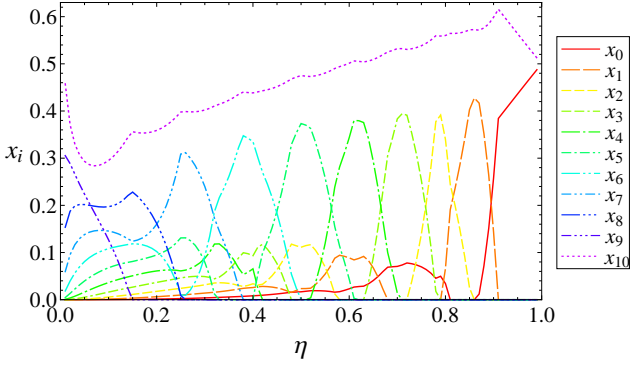}
\caption{%
  (color online). Parameters of the optimal state:
  $\ket{\psi}=\sum_{i=0}^N \sqrt{x_i} \ket{i,N-i}$, for phase
  estimation with $N=10$ photons for the case of losses in one only
  arm, i.e. $\eta_a=\eta$, $\eta_b=1$.}
\label{fig:optstates10}
\end{figure}

\begin{figure}
  \includegraphics[width=\columnwidth]{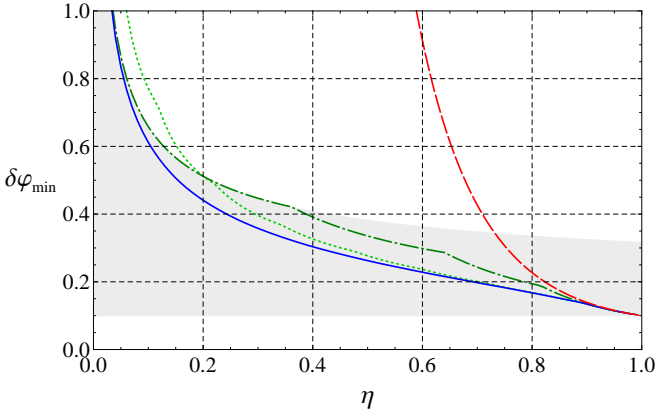}
\caption{%
  (color online). The minimal phase uncertainty achieved with various
  $N=10$ photon states for losses in one arm, i.e. $\eta_a=\eta$,
  $\eta_b=1$. Red, dashed line: N00N state; blue, solid line: optimal
  state (cf.  Fig.~\ref{fig:optstates10}); bright green, dotted line:
  optimal two-component state; dark green, dashed-dotted line: N00N
  chopping strategy. The shaded area is bounded by the SIL achievable
  with classical states and the Heisenberg limit $1/N$.}
\label{fig:optfisher10}
\end{figure}

The scenario of losses in only one arm of the interferometer, i.e.
$\eta_a=\eta$ and $\eta_b=1$, is relevant, for example, if losses are
induced by a sample itself. Figure~\ref{fig:optstates10} shows the
parameters $x_k$ of the optimal state for $N=10$ photons as a function
of $\eta$. The corresponding precision is shown in
Fig.~{\ref{fig:optfisher10}}. The shaded, grey area in
this figure is bounded by the SIL and the Heisenberg limit $1/N$, i.e.
whenever a line is in this region there is an improvement over
classical interferometry. The precision obtainable with the N00N state
 is worse than the SIL except for relatively low losses.

For transmissivities exceeding a certain threshold $\eta>\bar\eta$ the
optimal state is equal to the N00N state~(\ref{eq:NOON_state}). Note
that for $\eta=1$ the N00N state is ``balanced'', i.e. $x_0=x_N=1/2$.
We will give an estimate for the threshold $\bar\eta$ at the end of
this subsection. For $\eta<\bar\eta$ the optimal state consists of
more components. As seen in Fig.~\ref{fig:optfisher10} over a wide range we can achieve almost the
same precision by using two-component states of the form
\begin{equation}
\ket{\psi_{in}}= \sqrt{x_k} \ket{k,N-k} + \sqrt{x_N}\ket{N,0},
\end{equation}
where we numerically optimize $k$ and the amplitude $x_k$. With decreasing $\eta$ the
optimal choice for $k$ increases. This state is more robust in the
presence of losses than a N00N state, since a loss of up to $k$
photons in the first mode does not destroy the superposition, while
for the N00N state a loss of even a single photon renders the state
useless for phase estimation. On the other hand, a large $N-k$
increases the sensitivity of the state with respect to an induced
phase in channel $a$. The properties of large $k$ and large $N-k$ are
therefore competing and the optimal result represents a trade-off
between phase sensitivity and robustness.

An alternative strategy that leads to an improvement over the SIL, but
uses states of a simpler structure than the optimal state is the N00N
``chopping'' strategy which was introduced in Ref.~\cite{Dorner2008}.
Instead of a
single N00N state, we use the same number of photons, but send them
successively in $N/n$ smaller portions using $n$-photon n00n states.
Repeating an experiment $N/n$ times corresponds to an $N/n$ fold
increase in Fisher information, as shown in Eq.~(\ref{eq:cramer_repeat}).
Using Eq.~(\ref{eq:noon_prec}) we find that the Fisher information for
the N00N chopping strategy reads
\begin{equation}
F_Q= \frac{N}{n} \ \frac{4n^2 \eta^n}{(1+\eta^{n/2})^2}.
\end{equation}
Treating both $n$ and $N/n$ as real numbers with $n\in [1,N]$,
maximization of this expression over $n$ yields
\begin{equation}
\delta \varphi_{min}=
\begin{cases}
  \frac{1+\sqrt{\eta}}{2\sqrt{N\eta}}& ;\eta\le \eta_0 \approx 0.228 \\
  \frac{1+\sqrt{\eta_0}}{2\sqrt{N\eta_0}}\sqrt{\frac{\ln{\eta}}{\ln{\eta_0}}}&  ;\eta_0 <\eta\le \eta_0^{\frac{1}{N}}\\
  \frac{1+\eta^{\frac{N}{2}}}{2 N\eta^{\frac{N}{2}}}& ;\eta >
  \eta_0^{\frac{1}{N}}
\end{cases}
\end{equation}
where the optimal choice of $n$ for the three regimes indicated above
is $n=1$, $n=1.478/|\ln{\eta}|$ (i.e. the solution of $1+\eta^{n/2}+
n\ln\eta=0$) and $n=N$. We see that for very small transmissivities
$\eta\le \eta_0$ the precision is equal to the SIL,
specified in Eq.~(\ref{eq:sil}). For higher transmissivities the chopping strategy
beats the SIL, although only by a constant factor rather than in terms
of scaling.  For very high transmissivities $\eta>\eta_0^{1/N}$ the
best strategy is to use un-chopped N00N states. We note that a similar
strategy for multipartite qubit states has been devised
in~\cite{Shaji2007}.

As mentioned above the N00N state ceases to be optimal below a
threshold $\bar\eta$ in which case the coefficient $x_1$ obtains a
nonzero value. By adding an infinitesimal change $\delta x_1$ to $x_1$
at the expense of $x_0,x_N$ (which we assume are kept in the
proportion which is optimal in the absence of $x_1$), we can calculate
the corresponding change in the quantum Fisher information.
The change of $x_1$ by $\delta x_1$ increases $F_Q$ by $\frac{d F_Q}{d x_1} \delta x_1$,
but at the same time due to decreasing weight of the other coefficients decreases it by $F_Q \delta x_1$.
On the whole the change of $F_Q$ reads:
\begin{equation}
\delta F_Q= \left(\frac{d F_Q}{d x_1}-F_Q\right) \delta x_1.
\end{equation}
Substituting $F_Q$ from Eq.~(\ref{eq:fisher1mode}) at $x_0=\eta^{N/2}/(1+\eta^{N/2})$, $x_N=1-x_0$,
$x_{i}=0$ ($i=2,\dots,N-1$) and calculating $\delta F_Q$ we determine the value $\bar\eta$ below
which $\delta F_Q/\delta x_1$ is positive. This implies that for $\eta<\bar\eta$ an increase in $x_1$
will increase $F_Q$.  Writing $\delta F_Q /\delta x_1$ explicitly we get:
\begin{multline}
  \frac{\delta F_Q}{\delta x_1} = \frac{4}{(1+\eta^{N/2})^2}  [(1-2N) \eta^N  - 2N(N-1)\eta^{N/2+1} \\
  +2(N-1)^2 \eta^{N/2}- N(N-2) \eta +(N-1)^2].
\end{multline}
The roots of the expression in square brackets can be found
numerically. There is one real root in the interval $[0,1]$ which
corresponds to the threshold $\bar\eta$. The roots behave in very good
approximation like $\bar\eta=a^{-1/N}$ where $a\approx 2.61$ which is
obtained by a fit to the roots between $N=5$ and $N=100$. For example,
for $N=10$ we find that the threshold at which the N00N state ceases
to be optimal is at $\bar\eta=0.91$, which agrees with the numerical
results obtained before (see Fig.~\ref{fig:optstates10}).

\subsection{Equal losses in both arms}\label{lossesintwoarms}
In order to determine the optimal state in the case of equal losses in
both arms of the interferometer, i.e. $\eta_a=\eta_b\equiv\eta$, we
use $\tilde{F}_Q$ as given by Eq.~(\ref{eq:fisherbound}) as a basis
for our numerical optimization. It simplifies maximization drastically
since it is a concave function of the $\{x_k\}$. As pointed out in
Sec.~\ref{sec:calcfisherB}, $\tilde F_Q$, is only an upper bound to
the ``true'' quantum Fisher information $F_Q$ and therefore cannot be
reached by any measurement strategy if $F_Q$ is strictly smaller than
$\tilde F_Q$. However, using $\tilde{F}_Q$ can be still very useful:
We maximize $\tilde{F}_Q$ and use the resulting input state (say,
$\ket{\tilde\psi_{in}}$) to calculate the corresponding
$F_Q(\ket{\tilde\psi_{in}})$. On the one hand, the true maximum of
$F_Q(\ket{\psi_{in}})$ (corresponding to the truly optimal state,
$\ket{\psi_{in}}$) is certainly greater or equal to
$F_Q(\ket{\tilde\psi_{in}})$. On the other hand we have
$F_Q(\ket{\psi_{in}}) \leq \tilde{F}_Q(\ket{\psi_{in}})$ due to
convexity [see Eq.~(\ref{eq:boundconvexity})]. But since
$\tilde{F}_Q(\ket{\tilde\psi_{in}})$ is the global maximum we have
$\tilde{F}_Q(\ket{\psi_{in}}) \leq
\tilde{F}_Q(\ket{\tilde\psi_{in}})$.  Consequently, the true maximum
must lie \emph{in between} $F_Q(\ket{\tilde\psi_{in}})$ and
$\tilde{F}_Q(\ket{\tilde\psi_{in}})$. Hence, if
$|\tilde{F}_Q(\ket{\tilde\psi_{in}})- F_Q(\ket{\tilde\psi_{in}})|$ is
small on a scale which is for this problem naturally given by the
difference between the SIL and the Heisenberg limit, we obtain a very
good approximation of the true maximal value of the quantum Fisher
information. We showed numerically that for the examples discussed in
this section this difference never exceeds $0.4\%$.
\begin{figure}
  \includegraphics[width=\columnwidth]{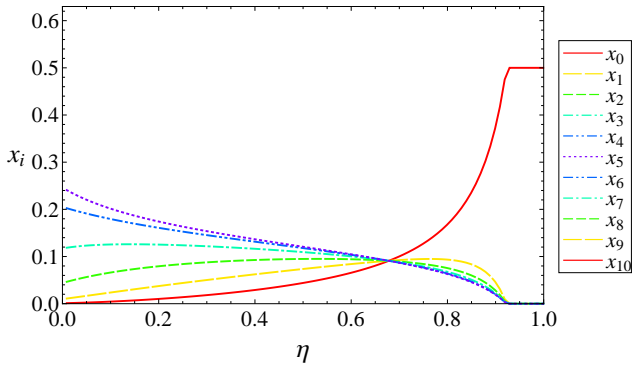}
\caption{%
  Parameters of the optimal state: $\ket{\psi}=\sum_{i=0}^N
  \sqrt{x_i}\ket{i,N-i}$, for phase estimation with $N=10$ photons for
  the case of equal losses in both arms, i.e.  $\eta_a=\eta_b=\eta$.
  Due to this symmetry we have $x_k=x_{N-k}$.}
\label{fig:optstates2modes10}
\end{figure}
\begin{figure}
  \includegraphics[width=\columnwidth]{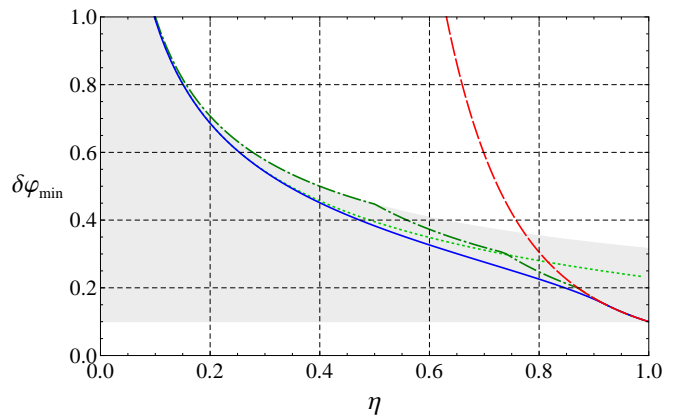}
\caption{%
  The minimal phase uncertainty achieved with various $N=10$ photon
  states for equal losses in both arms, i.e. $\eta_a=\eta_b=\eta$.
  Red, dashed line: N00N state; blue, solid line: optimal state (cf.
  Fig.~\ref{fig:optstates2modes10}); bright green, dotted line: the
  state optimal for lossless phase estimation in the global approach
  [see Eq.~(\ref{eq:stateoptglobal})]; dark green, dashed-dotted
  line: the N00N chopping strategy. The shaded area is bounded by the SIL
  achievable with classical states and the Heisenberg limit $1/N$.}
\label{fig:optfisher2modes10}
\end{figure}

The result of optimization is shown in
Fig.~\ref{fig:optstates2modes10} for $N=10$. Due to the symmetry of
the problem, $\eta_a=\eta_b$, the solution has the property $x_k =
x_{N-k}$.  For small losses the optimal state is the balanced N00N
state, i.e. $x_0=x_N=1/2$. However, below a threshold $\bar\eta$ all
other terms $x_k$ obtain non-vanishing values. With decreasing $\eta$,
$x_0$ and $x_N$ get smaller, and eventually the coefficients $x_k$
with $k\approx N/2$ are dominant.  Interestingly, this structure is
qualitatively similar to a two-mode generalization of the state which
is optimal for one-mode lossless phase estimation in the global
approach \cite{Berry2000}
\begin{equation}
\label{eq:stateoptglobal}
\ket{\psi_{in}}=\sqrt{2/(N+2)}\sum_{k=0}^N \sin[\pi(k+1)/(N+2)]\ket{k,N-k}.
\end{equation}
It was noted in Ref.~\cite{Maccone2008} that these states perform
quite well in the presence of loss, and hence could be useful for the
problem we consider.

The threshold $\bar\eta$ can be obtained in a similar fashion as in
the previous subsection. By adding an infinitesimal change $\delta
x_k$ to a component $x_k$ other than $x_0$ and $x_N$ we can derive a
polynomial which has a root in $[0,1]$ defining $\bar\eta(N)$. Since
the polynomial is rather cumbersome we do not quote it here, however,
the relevant root is again given in very good approximation by
$\bar\eta = a^{-1/N}$ where $a\approx 2.24$ is obtained by a fit to
the roots between $N=2$ and $N=100$. For example for $N=10$
(corresponding to Figs.~\ref{fig:optstates2modes10}
and~\ref{fig:optfisher2modes10}) we get $\bar\eta = 0.92$.

Analogous to the case of losses in only one arm we can examine a N00N
chopping strategy by sending ``smaller'' n00n states through the
interferometer. The corresponding precision reads
\begin{equation}
\delta\varphi_{min} =
\begin{cases}
  \frac{1}{\sqrt{N\eta}} & ;\eta\le \e^{-1} \\
  \sqrt{\frac{\e|\ln{\eta}|}{N}} & ;\e^{-1}<\eta\le \e^{-\frac{1}{N}}\\
  \frac{1}{N\eta^{\frac{N}{2}}} & ;\eta >
  \e^{-\frac{1}{N}}.
\end{cases}
\label{eq:chop1}
\end{equation}

Figure~\ref{fig:optstates2modes10} shows the estimation precision for
the optimal state (blue line; here we use
$\delta\varphi_{min}=1/\sqrt{\tilde F_Q}$), the N00N state (red line),
the N00N chopping strategy (dark green line), and the
state~(\ref{eq:stateoptglobal}) (bright green line). In the latter
case the precision is indeed quite high in the regime of high losses,
outperforming the N00N chopping strategy but for small losses it is
significantly worse than both the optimal and the N00N chopping
strategies \footnote{Claims in \cite{Maccone2008}, that the state
  given by Eq.~(\ref{eq:stateoptglobal}), outperforms even the optimal
  states we have presented, arise from taking a different perspective
  in quantifying the resources. See Sec.~\ref{sec:indefiniteN} for
  more discussion.}.

\section{The optimal measurement}
\label{sec:measurement}
As discussed in Sec.~\ref{sec:fisher}, a measurement saturating the
quantum Cram\'er-Rao bound always exists.  For the cases when the
states $\ket{\xi_{l_al_b}(\varphi)}$ introduced in
Sec.~\ref{sec:interferometry} are orthogonal for different $l=l_a+l_b$
(i.e. when $F_Q=\tilde{F}_Q$) we can derive an explicit POVM
saturating the bound. To this end we first assume that we perform a
measurement determining the total number of photons lost, $l$, which
projects the system onto a particular pure state
$\ket{\xi_{l_al_b}(\varphi)}$ \footnote{Up to now, no effective
  non-destructive photon number measurement was realized.
  Nevertheless, as most quantum optical experiments work in the
  post-selection paradigm, this is not really a limitation, since the
  number of photons can be detected afterwards, and only appropriate
  events taken into account.}.  Subsequently, we perform the
measurement saturating the Cram\'er-Rao bound on pure states (see
Sec.~\ref{sec:fisher}), i.e. a projection on a basis containing the
vectors
\begin{multline}
  \ket{e^{l_a l_b}_{\pm}}= \frac{1}{\sqrt{2}}\left( \ket{\xi_{l_al_b}(\varphi)} \pm \ket{\xi^{\prime \bot}_{l_a l_b}(\varphi)} \right) = \\
=\frac{1}{\sqrt{2}}  \sum_{k=l_a}^{N-l_b} \alpha_k \sqrt{B^k_{l_al_b}} \left(1\pm \frac{i
      (k-l_a -\langle \hat{n}_a \rangle_{l_al_b})
    }{\sqrt{(\Delta \hat{n}_a)^2_{l_al_b}}} \right)e^{i k \varphi}
  \ket{k-l_a, N-k-l_b},
\end{multline}
where $\langle \hat{n}_a \rangle_{l_al_b}$,
$(\Delta \hat{n}_a)^2_{l_al_b}$ are the mean photon number and the
variance of the photon number in mode $a$ for the state
$\ket{\xi_{l_al_b}(\varphi)}$.
The above POVM depends, in general, on the phase
$\varphi$.

There are cases in which a single POVM saturating the Cram\'{e}r-Rao bound for all values of $\varphi$ can be found.
In a single photon case ($N=1$), states of the form $1/\sqrt{2}(\ket{0,1}+e^{i \chi}\ket{1,0})$, allow for a $\varphi$-independent POVM,
whereas states with unbalanced weights of $\ket{01}$, and $\ket{10}$ terms lead to optimal POVMs which are necessarily $\varphi$-dependent.
Moreover, for an $N$ photon state $\ket{\psi}= \sum_{k=0}^N \alpha_k \ket{k,N-k} $, a single, $\varphi$-independent
POVM saturating the Cram\'{e}r-Rao bound can be found provided that the state enjoys
the path-symmetry property: $\alpha_{k}=\alpha^*_{N-k}e^{i \chi}$, where $\chi$ is a fixed phase the same for all components \cite{Hofmann2009}.
In general, due to losses the conditional pure states that appear in our problem lack the above mentioned symmetry and
 consequently the optimal POVM varies with the change of $\varphi$.
\section{More general input states}
\label{sec:generalstates}
One could argue that the $N$ photon state given by
Eq.~(\ref{eq:input}) does not represent the most general way of using
$N$ photons to determine the relative phase in an interferometer:
Instead of considering a state with a \emph{definite} photon number
$N$, one could consider a state which is a superposition of terms with
different total photon number, and only imply a constraint that the
\emph{mean} photon number is $N$. Furthermore, by using states of the
form~(\ref{eq:input}) it is assumed that all photons in an arm of the
interferometer are indistinguishable, i.e.  occupy a common
spatio-temporal mode. This greatly limits the class of considered
states since the Hilbert space dimension of states sent into the
interferometer would be of the order $2^N$ if the photons were
distinguishable, while for indistinguishable photons it is $N+1$. In
the next two subsections we show that none of these generalizations
improves the phase estimation precision, and consequently it is
sufficient to consider the states of the form given by
Eq.~(\ref{eq:input}).

\subsection{States with indefinite photon number}
\label{sec:indefiniteN}
We emphasize that in this paper we consider closed
systems, i.e.\ there are no additional reference beams, neither classical nor
quantum, since any additional beam would contain photons and should therefore
be explicitly taken into account for the determination of the required resources.

Consider a superposition of input states of the interferometer
with different, but definite photon numbers $n$,
\begin{equation}
\label{eq:superpos}
\ket{\Psi_{in}}=\sum_{n=0}^{\infty} \beta_n \ket{\psi_{in}^{(n)}},
\end{equation}
where the mean total photon number is $N=\sum_{n=0}^\infty
|\beta_n|^2 n $.  Under free evolution a
term with $n$ photons acquires a phase $e^{-i n \omega t}$, i.e.
terms with different $n$ evolve with different
frequencies.  In the absence of an additional reference beam which
allows for clock synchronization between the sender and the receiver
the relative phases between terms with different $n$
become unobservable and the state given by Eq.~(\ref{eq:superpos}) is
physically equivalent to a mixture \cite{Molmer1997, Bartlett2006},
\begin{equation}
\rho_{in}=
 \sum_{n=0}^\infty |\beta_{n}|^2 \ket{\psi^{(n)}_{in}} \bra{\psi^{(n)}_{in}}.
\end{equation}
Moreover, since the quantum Fisher information is convex (see
Sec.~\ref{sec:fisher}), we have
\begin{equation}
F_Q(\rho_{in}) \leq \sum_{n=0}^\infty |\beta_{n}|^2 F_Q(\ket{\psi^{(n)}_{in}}\bra{\psi^{(n)}_{in}}).
\end{equation}
Consequently it is always better to send a state with a fixed number
of photons $\ket{\psi_{in}^{(n)}}$, with probability $|\beta_n|^2$,
rather than to use a superposition (which is effectively a mixture).
The analysis can thus be restricted to states with definite photon
number without compromising optimality.

\subsection{Distinguishability of photons}
In this subsection we consider general $N$ photon input states where
the photons are distinguishable which is the case, e.g., if they are
sent in different time bins. A state of this type can be written as
\begin{equation}
\ket{\psi_{in}}=\sum_{\boldsymbol{k}=\boldsymbol{0}}^{\boldsymbol{1}} \alpha_{\boldsymbol{k}}\ket{ \boldsymbol{k}},
\end{equation}
where $\boldsymbol{k}$ represents a binary sequence of length $N$, and
$\boldsymbol{0}\equiv\{0,\dots,0\}$,
$\boldsymbol{1}\equiv\{1,\dots,1\}$.  Summation from a sequence to a
sequence should be understood using partial ordering between
sequences, i.e. $\boldsymbol{k} \leq \boldsymbol{k}^\prime
\Leftrightarrow k_i \leq k^\prime_i \,\,\forall i$.  Therefore the
summation is performed over all binary sequences. Furthermore
$\ket{\boldsymbol{k}}=\ket{k_1}\otimes \dots \otimes \ket{k_N}$
denotes a state of $N$ photons, where $k_i=0 (1)$ means that the
photon sent in the $i$-th time bin propagates in arm $a$ ($b$) of the
interferometer.  Mathematically, we deal here with an arbitrary
$N$-qubit state, which lives in a $2^N$ dimensional space.  Notice,
that the indistinguishable case is recovered once we consider only the
fully symmetric (bosonic) subspace of $N$ qubits.

$\tilde{F}_Q$ can be calculated in a similar way as in
Sec.~\ref{sec:calcfisher} and reads
\begin{equation}
\label{eq:fisherSM}
\tilde{F}_Q=4 \left(\sum_{\boldsymbol{k}=\boldsymbol{0}}^{\boldsymbol{1}}
\bar{\boldsymbol{k}}^2 x_{\boldsymbol{k}} -
\sum_{\boldsymbol{l}_a=\boldsymbol{0}}^{\boldsymbol{1}} \sum_{\boldsymbol{l}_b=0}^{\boldsymbol{1}-\boldsymbol{l}_a}
\frac{\left(\sum_{\boldsymbol{k}=\boldsymbol{l}_a}^{\boldsymbol{1}-\boldsymbol{l}_b}
\bar{\boldsymbol{k}} x_{\boldsymbol{k}}
B^{\boldsymbol{k}}_{\boldsymbol{l}_a\boldsymbol{l}_b}\right)^2}{\sum_{\boldsymbol{k}=\boldsymbol{l}_a}^{\boldsymbol{1}-\boldsymbol{l}_b}
x_{\boldsymbol{k}} B^{\boldsymbol{k}}_{\boldsymbol{l}_a\boldsymbol{l}_b}}
\right),
\end{equation}
where boldface symbols denote binary strings of length $N$,
$x_{\boldsymbol{k}}=|\alpha_{\boldsymbol{k}}|^2$,
$\bar{\boldsymbol{k}}$ denotes the number of $1$s in the sequence
$\boldsymbol{k}$, and
\begin{equation}
B^{\boldsymbol{k}}_{\boldsymbol{l_a}\boldsymbol{l_b}}=(1-\eta_a)^{\bar{\boldsymbol{l}}_a} \eta_a^{\bar{\boldsymbol{k}}-\bar{\boldsymbol{l}}_a}
(1-\eta_b)^{\bar{\boldsymbol{l}}_b} \eta_b^{N-\bar{\boldsymbol{k}}-\bar{\boldsymbol{l}}_b}.
\end{equation}
Again $\tilde{F}_Q = F_Q$ provided that knowledge from which mode a photon was lost is not relevant or self-evident, e.g. when $\eta_b=1$.
Let us now symmetrize the state of $N$ photons. This corresponds to
replacing $x_{\boldsymbol{k}}$ with $1/N!\sum_\sigma
x_{\sigma(\boldsymbol{k})}$, where the summation is performed over all
permutations of an $N$ element set. Notice that the first term within
the parenthesis in Eq.~(\ref{eq:fisherSM}) is not affected by
symmetrization. Let us denote
$a_{\boldsymbol{l}_a\boldsymbol{l}_b}=\sum_{\boldsymbol{k}=\boldsymbol{l}_a}^{\boldsymbol{1}-\boldsymbol{l}_b}
\bar{\boldsymbol{k}} x_{\boldsymbol{k}}
B^{\boldsymbol{k}}_{\boldsymbol{l}_a\boldsymbol{l}_b}$ and
$b_{\boldsymbol{l}_a\boldsymbol{l}_b}=\sum_{\boldsymbol{k}=\boldsymbol{l}_a}^{\boldsymbol{1}-\boldsymbol{l}_b}
x_{\boldsymbol{k}}
B^{\boldsymbol{k}}_{\boldsymbol{l}_a\boldsymbol{l}_b}$. Then the
subtracted term in Eq.~(\ref{eq:fisherSM}) reads
$\sum_{\boldsymbol{l}_a=\boldsymbol{0}}^{\boldsymbol{1}}
\sum_{\boldsymbol{l}_b=0}^{\boldsymbol{1}-\boldsymbol{l}_a}
\frac{a^2_{\boldsymbol{l}_a\boldsymbol{l}_b}}{b_{\boldsymbol{l}_a\boldsymbol{l}_b}}$.
Performing symmetrization corresponds to replacing
$a_{\boldsymbol{l}_a\boldsymbol{l}_b}$
and $b_{\boldsymbol{l}_a\boldsymbol{l}_b}$ with $1/N! \sum_{\sigma}
a_{\sigma(\boldsymbol{l}_a),\sigma(\boldsymbol{l}_b)}$ and $1/N!
\sum_{\sigma} b_{\sigma(\boldsymbol{l}_a),\sigma(\boldsymbol{l}_b)}$ respectively.
Since $b_{\boldsymbol{l}_a\boldsymbol{l}_b}$ are positive, the
following inequality holds
\begin{equation}
\frac{a_{\boldsymbol{l}_a\boldsymbol{l}_b}^2}{b_{\boldsymbol{l}_a\boldsymbol{l}_b}} +
\frac{a_{\boldsymbol{l}^\prime_a\boldsymbol{l}^\prime_b}^2}{b_{\boldsymbol{l}^\prime_a\boldsymbol{l}^\prime_b}}
 \geq
2 \frac{[1/2(a_{\boldsymbol{l}_a\boldsymbol{l}_b}+a_{\boldsymbol{l}^\prime_a\boldsymbol{l}^\prime_b})]^2}{1/2(b_{\boldsymbol{l}_a\boldsymbol{l}_b}+
b_{\boldsymbol{l}^\prime_a\boldsymbol{l}^\prime_b})},
\end{equation}
which shows that symmetrizing can only decrease the subtracted
fraction, hence increase the Fisher information.  This proves that the
optimal states are symmetric states living in the bosonic subspace.
Consequently, using indistinguishable photons is sufficient to obtain
the optimal Fisher information.

A physical intuition behind the above derivation is the following.
Notice that if there are no losses, the optimal $N$ qubit state has
the form $1/\sqrt{2}(\ket{\boldsymbol{0}} + \ket{\boldsymbol{1}})$ --
it is the N00N state which lives in the fully symmetric subspace. In
this case there is no advantage in using distinguishable photons.  If
there are losses, things get even worse for distinguishable photons.
There is still no advantage in terms of phase sensitivity, yet when a
photon is lost, the \emph{knowledge of which photon was lost}
additionally harms the quantum superposition. Hence, it is optimal to
use states from a fully symmetric subspace, where it is not possible to
tell which photon was actually lost.

\section{Summary}
We have analyzed the optimal way of using $N$-photon states for
phase interferometry in the presence of losses.  We have derived an
explicit formula for the quantum Fisher information in the case when
losses are present in one arm of the interferometer, and have provided a
useful bound for the case of losses in both arms.  Using the quantum
Fisher information as a figure of merit, we have found the optimal
states, investigated their advantage over various quantum and
classical strategies. The optimal measurement saturating the Cram{\'e}r-Rao bound
has been presented, and it has been proven that in general it
varies depending on the phase in the vicinity of which we perform estimation.

A close inspection of the properties of the quantum Fisher information showed that it is optimal to use a quantum state with a definite
number of indistinguishable photons. Neither allowing superpositions of
different total photon-number terms, nor making photons distinguishable can
improve estimation precision.

\acknowledgments
This research was supported by the EPSRC (UK) through the QIP IRC
(GR/S82176/01), the AFOSR through the EOARD, the European Commission
under the Integrated Project QAP (Contract No.~015848), the Royal
Society and the Polish MNiSW (N N202 1489 33).


\end{document}